# Physics-based reward driven image analysis in microscopy

K. Barakati,[1,a] Hui Yuan,[3] Amit Goyal,[4] and S. V. Kalinin,[1,2, b]

[1]*Department of Materials Science and Engineering, University of Tennessee, Knoxville, TN 37996, USA*

[2]*Pacific Northwest National Laboratory, Richland, WA 99354, USA*

[3]*Department of Materials Science and Engineering, McMaster University, 1280 Main Street West, Hamilton, Ontario, L8S 4L7 Canada*

[4]*Laboratory for Heteroepitaxial Growth of Functional Materials & Devices, Department of Chemical & Biological Engineering, State University of New York, Buffalo, NY, 14260, USA*

**ABSTRACT:**

The rise of electron microscopy has expanded our ability to acquire nanometer and atomically resolved images of complex materials. The resulting vast datasets are typically analyzed by human operators, an intrinsically challenging process due to the multiple possible analysis steps and the corresponding need to build and optimize complex analysis workflows. We present a methodology based on the concept of a Reward Function coupled with Bayesian Optimization, to optimize image analysis workflows dynamically. The Reward Function is engineered to closely align with the experimental objectives and broader context and is quantifiable upon completion of the analysis. Here, cross-section, high-angle annular dark field (HAADF) images of ion-irradiated (Y, Dy)$Ba_2Cu_3O_{7-\delta}$ thin-films were used as a model system. The reward functions were formed based on the expected materials density and atomic spacings and used to drive multi-objective optimization of the classical Laplacian-of-Gaussian (LoG) method. These results can be benchmarked against the DCNN segmentation. This optimized LoG* compares favorably against DCNN in the presence of the additional noise. We further extend the reward function approach towards the identification of partially-disordered regions, creating a physics-driven reward function and action space of high-dimensional clustering. We pose that with correct definition, the reward function approach allows real-time optimization of complex analysis workflows at much higher speeds and lower computational costs than classical DCNN-based inference, ensuring the attainment of results that are both precise and aligned with the human-defined objectives.

[a] K.barakat@vols.utk.edu
[b] sergei2@utk.edu

Electron and scanning probe microscopy have emerged as a primary method to provide insights into the microstructure, composition, and properties of a wide range of materials, from metals and alloys to polymers and composites.[1-4] These techniques generate large volumes of imaging data containing information on material structure that can be further connected to fundamental physics and chemistry, processing, etc.[5] However, the large amount of imaging data requires consistent analysis methods.[6-8] Traditionally, this has been accomplished using the collection of the standard image processing techniques including various forms of background subtraction[9,10], filtering[11], and peak finding[12-14], all applied by the human operator sequentially.[15] The employment of machine learning methodologies, particularly DCNN (Deep Convolutional Neural Network) segmentation[16-20], has notably enhanced and expedited certain steps within this analytical framework; however, the overall progression of image analysis remains the same. This type of analysis is also computationally intensive[21-24] and requires the ensemble networks to effectively manage deviations from anticipated data distributions.[25] Most importantly, it is strongly biased by the operator's expertise and can potentially be steered towards anticipated answers via decisions made at each analysis step.

Here we present a method for image analysis that utilizes a reward function concept [26,27]. This involves setting a measure(s) of success that can be quantitatively established by the end of the analysis. With the reward function defined, the analysis workflow including the sequence and hyper-parameters of individual operations can be optimized via one of the suitable stochastic optimization frameworks. The simple image analysis workflow is optimized by Bayesian Optimization[28-31] which allows dynamic tuning of the parameters to achieve optimal performance. This concept can be further adapted to more complex, multi-stage workflows via reinforcement learning, Monte Carlo decision trees, or more complex algorithms.[32,33]

In proposing reward-driven workflows, we note that typically human-based image analysis is performed to optimize certain implicit measures of the analysis quality. For example, in atomic segmentation, this task is to identify and classify all the atoms of a certain type, or all defects within the image. Here we propose that analysis can be cast as an optimization problem if the reward function based on the analysis results can be formulated. Then the process becomes optimized in the parameter space of the simple analysis functions. Here, we consider two specific tasks, namely atom finding in atomically resolved images and identification of amorphized regions within the material.

As a model system, we chose a 1.2 μm thick $YBa_2Cu_3O_{7-\delta}$ thick film, doped with $Dy_2O_3$ nanoparticles, fabricated using a metal-organic deposition process. The sample then was irradiated with an $Au^{5+}$ ion beam oriented along the c-axis of the Yttrium Barium Copper Oxide (YBCO), and the cross-sectional and plan-view TEM specimens were prepared through standard mechanical polishing, followed by final thinning using Xe Plasma Focused Ion Beam (Xe PFIB).[34]

As a first model task, we consider the semantic segmentation,[35-38] or "atom finding" of atomically resolved images.[39] Traditionally this has been accomplished using the peak finding procedures, correlative filtering, Hough transforms[40,41], or versions of Laplacian of Gaussian (LoG) approaches[42,43]. These approaches require extensive tuning of the parameters of the image analysis function with the human assessment of the results as feedback. The introduction of DCNNs has resulted in broad interest in deep learning segmentation of images[44-46], with multiple efforts utilizing versions of U-Nets[47,48], masked RCNNs[49], and other architectures reported recently. The use of simple analysis methods requires careful manual tuning of parameters and tends to be very brittle – the contrast variations even within a single image can result in measurable differences of performance. Comparatively, DCNN methods are more robust, but require supervised training and can be sensitive to out-of-distribution drift effects.[50-52]

Taking atom detection as an initial instance of the reward-driven process, we demonstrate optimization of the conventional Laplacian of Gaussian (LoG) algorithm. This approach is characterized by a set of control parameters including min_sigma ($\sigma_{min}$), max_sigma ($\sigma_{max}$), num_sigma ($\sigma_{num}$), threshold (T), and overlap ($\theta$), which define its parameter space, as illustrated in **Figure 1(A)**.

To cast the image analysis as an optimization problem, we define possible physics-based reward (or objective) functions. One such function can be defined based on the expected number of atoms within the field of view, readily available from image size and lattice parameter of material. The LoG algorithm's effectiveness in relation to its hyper-parameters is determined by a metric we refer to as Quality Count (QC), which is defined as the normalized difference between the number of atoms found by Laplacian of Gaussian (LoG) and the physics-based reward standard, formulated as:

$$QC = \frac{LoG\ blobs - Physics\_blobs}{Physics\_blobs} \qquad (1)$$

To avoid reward hacking in this context, we also recognize that the total count of atoms is an overarching characteristic, and for a segmentation algorithm to be effective, it should adhere to more specific requirements. The second constraint is that atoms need to be spaced at distances that are physically plausible. To incorporate this aspect, we introduce a second component to the reward function, which we call the Error function.

The Error function (ER) will be applied to measure the incidence of atoms in regions that are not aligned with the structural configuration of the YBCO lattice. As shown in **Figure 1(B)**, any atom with a cumulative interaction value less than DS (summation of distances) with its four nearest neighbors will be regarded as not having physical significance, and thus, categorized as an error within this context.

$$ER = \frac{\#\ atoms\ with\ cumulative\ interaction\ value\ less\ than\ DS}{Physics\_Blobs} \qquad (2)$$

In this setting, the optimization of LoG analysis that we will further refer to as LoG*, becomes that of the multi-objective Bayesian Optimization in the image processing parameter space, where objectives QC and ER are minimized jointly.
In this case, we can further define a benchmark for accuracy, which we designate as "Oracle" in this context. A possible way to create an Oracle for the atomic segmentation task can be performed using the pre-trained DCNN, providing near-ideal identification of all atomic positions. These can be further classified (with human tuning) into specific types. We refer to the DCNN analysis as "Oracle" comparable to human-based analysis and use Oracle to verify the results of the reward-driven workflows accomplished with much simpler tools.

We employed the *Skopt* library[53, 54] to implement hyper-parameter optimization, specifically focusing on adjusting the threshold and overlap parameters of the Laplacian of Gaussian (LoG) function. As represented in **Figure 1(C)**, a set of optimal solutions, or Pareto front, where no objective can be improved without degrading the other was obtained. Through this framework, a delicate balance between the dual objectives has been established, leading to the discovery of an optimal hyper-parameter configuration for the LoG function. Two common metrics to identify the "best" solutions within the Pareto Frontier are the Euclidean and Chebyshev distances.

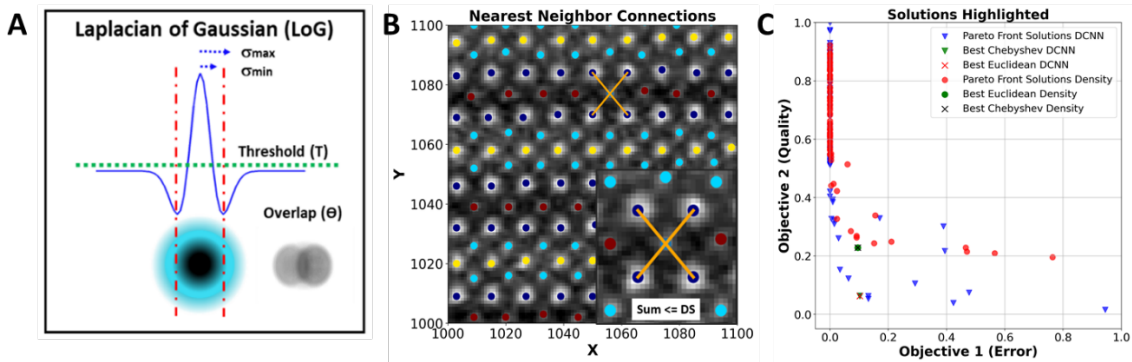

**Figure 1:** A) Laplacian of Gaussian Hyper-parameters, min_sigma ($\sigma_{min}$), max_sigma ($\sigma_{max}$), num_sigma ($\sigma_{num}$), threshold (T), and overlap ($\theta$), B) Error function definition based on the presence of atoms in areas that deviate from the structural arrangement of the YBCO lattice, C) Pareto Frontier solutions with respect to Oracle-A, and Oracle-B). Each point represents an optimal trade-off point such that improving one objective would compromise another. This balance delineates the optimal hyper-parameter settings for the Laplacian of Gaussian (LoG) function, achieved by finely tuning the competing objectives.

Displayed in **Figure 2(A)** is the workflow development utilized for Multi Objective-Bayesian Optimization. This workflow outlines the order of steps throughout the analysis procedure. We note that this approach can be readily applied to the scenarios when the image quality or acquisition conditions vary across the image, e.g., due to the mis-tilt or presence of non-crystalline contaminates, etc. For these tasks, the algorithm can be implemented in the sliding window setting where the parameters are optimized for each. Further, this workflow can be customized to focus on different rewards such as the identification of the amorphous regions or other objectives of the study as presented in **Figure 2(B)**.

As a next step, we explore the robustness of the proposed approach with respect to the noise in the image. To accomplish this, Gaussian noise levels from 0 to 1, where 0 is the image without noise have been applied to a specific set of images. Upon noise addition, the number of atoms is identified both by DCNN and optimized (LoG*) algorithm. **Figure 3(A)** depicts the variation in optimal hyperparameters of the LoG model in response to different levels of added noise. Correspondingly, **Figure 3(D)** demonstrates that the best Pareto front solutions, which represent the objectives (QC and ER), adapt in a manner that fulfills the reward requirements.

In DCNN models, elevating the noise level leads to the introduction of artifacts that mimic the appearance of new atoms in the images, thereby generating false positives as depicted in **Figure 3(C)**. In contrast, the LoG function demonstrates resilience when subjected to comparable increases in noise, avoiding the misidentification of these artifacts as new atoms, as evidenced in **Figure 3(B)**. This stability can be attributed to the implementation of the ER function within the LoG framework, which effectively prevents the function from mistakenly identifying features caused by noise as real atomic points.

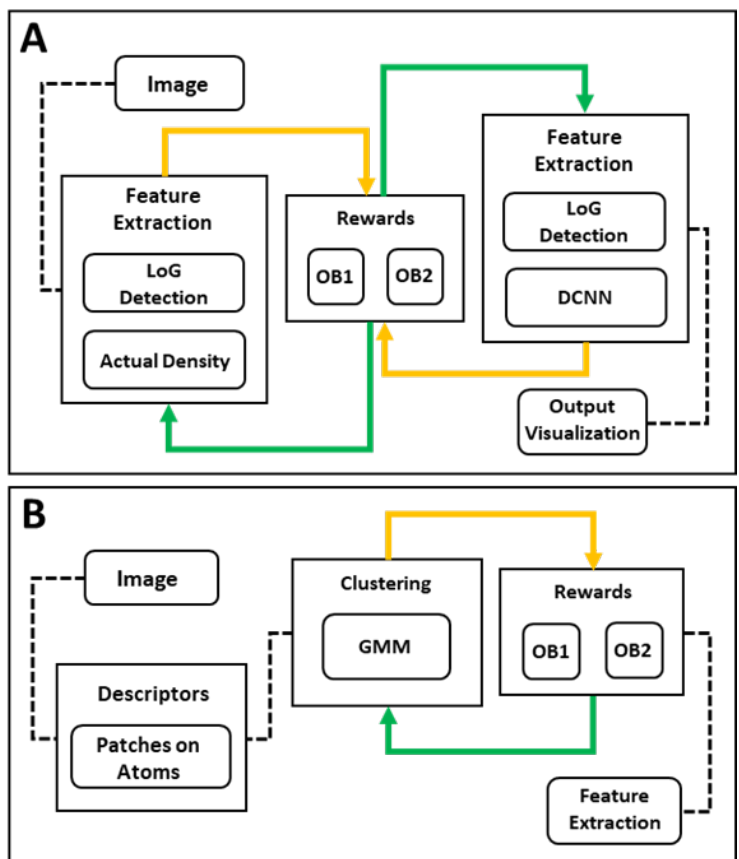

**Figure 2:** A) Workflow for reward-driven methodology in TEM images, Data preparation, Optimization of the LoG function based on two objectives using Multi-objective Bayesian Optimization, and Processing, B) Workflow for reward-driven methodology in TEM images, task specified version.

    **Figure 3(F)** illustrates the detection capability of the DCNN model regarding Gaussian noise levels. The number of detected atoms increases significantly with the Gaussian noise level after a certain point (Noise level of 0.6), which implies that the DCNN begins to mistakenly identify noise artifacts as atoms, thereby detecting false positives. **Figure 3(E)** represents the detection results of the LoG method under the same conditions. In contrast to the DCNN, the LoG detection exhibits a much lower variability in the number of detected atoms across noise levels, maintaining a relatively consistent count. This implies that the LoG approach is more selective, mainly identifying actual atomic points and not creating false positives by noise-related distortions.

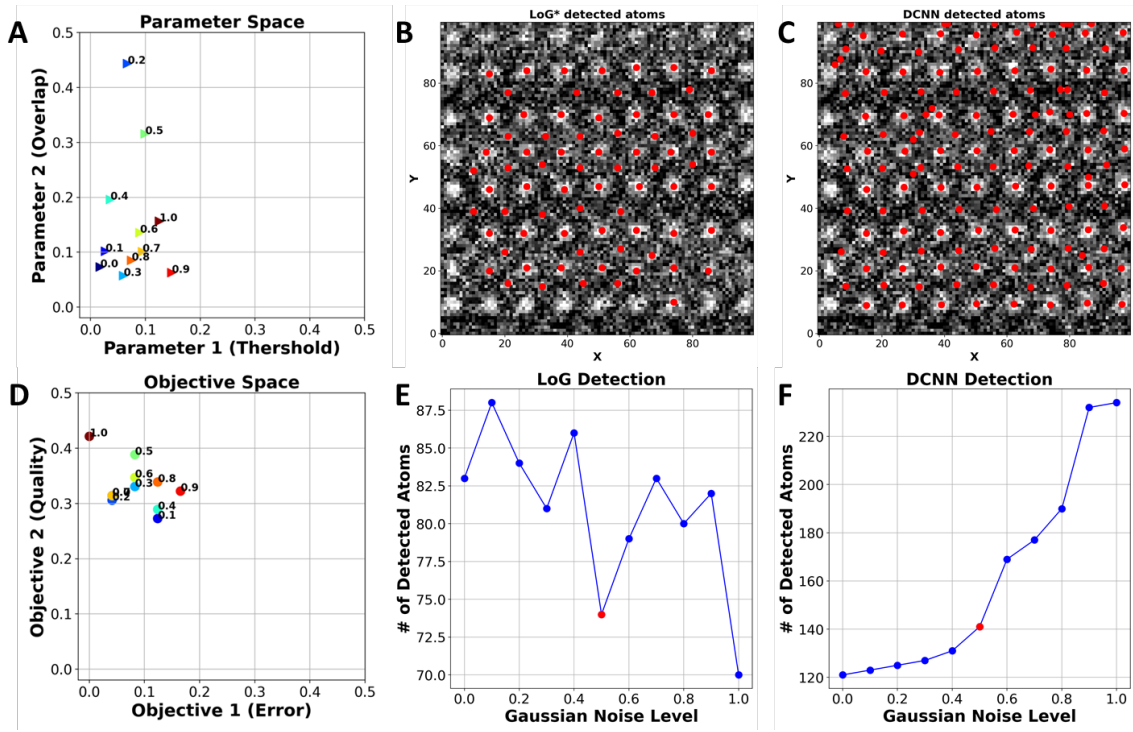

**Figure 3:** A) Optimal hyper-parameter space changes verses noise level in the LoG* optimized method, B) Detected atoms using the LoG* optimized method on the image with a moderate noise level, C) Detected atoms using a (DCNN) model scattered on the image, D) Optimal objective space change verses noise level in the LoG* optimized method, E) Number of detected atoms versus Gaussian noise level using the LoG* optimized method, F) Number of detected atoms versus Gaussian noise level using a (DCNN) model.

We have further explored the applicability of this approach towards more complex tasks of identification of the amorphous regions. Here, the complexity of analysis is that the damage introduces amorphization and change of observed image contrast on oxygen and copper lattices, whereas the bright atoms remain visible. Correspondingly, manual construction of the workflow combining segmentation, multiple possible clustering and dimensionality reduction algorithms can be a very time-consuming and operator-dependent step. Here we illustrate that the use of the reward function approach allows us to solve this problem via workflow combining window size selection and automated parameter tuning of a specific clustering method, namely the Gaussian Mixture Model (GMM).

Considering the workflow in **Figure 2(B)**, initially, we implemented GMM clustering techniques to identify the diverse atomic configurations within the YBCO structure. **Figure 4(A)** displays the categorization of all atomic types present in the YBCO structure. We organized these into four distinct clusters corresponding to the $CuO_2$ (Planes), CuO (Chains), Ba (Barium), and Y (Yttrium) components, respectively. Given that certain atomic varieties can dominate the clustering outcomes, we refined our approach by reducing the number of cluster types to specifically focus on barium (Ba) atoms. This was achieved by conducting two separate (GMM) clustering analyses on patches centered exclusively on Ba atoms. As illustrated in **Figure 4(B),** two distinct clusters were identified, corresponding to the orientation of barium (Ba) atoms. These clusters are categorized based on their orientation: $Ba_1$ is aligned along a principal axis, while $Ba_2$ is configured to exhibit two-fold rotational symmetry with respect to $Ba_1$. By concentrating solely on $Ba_1$ or $Ba_2$ atoms, (GMM) clustering enables us to detect the variations in Ba atoms.

In this context, following the logic that atoms in crystalline regions are well-organized and show minimal deviations from their expected lattice positions, resulting in a tightly packed clustering. However, any observed dispersity within these clusters serves as a clear indicator of deviations from the expected lattice positions, which is characteristic of atoms in amorphous areas. This distinction allows for the differentiation of crystalline and amorphous structures based on the spatial arrangement and variability of atomic positions.

**Figure 4(C)** demonstrates that the clustering of atomic points can be controlled through the adjustment of two hyper-parameters of GMM: threshold and covariance type. According to our hypothesis, atomic points that surpass a predetermined threshold, when analyzed using a specific covariance type, should be classified as amorphous. This classification is substantiated by the observed dispersity of these points away from the core cluster, which is predominantly associated with crystalline regions. In this instance, the effectiveness of (GMM) clustering depends primarily on hyper-parameter selection and can be improved by devising a customized reward system that better aligns with desired outcomes.

To direct (GMM) clustering toward not only pinpointing the location but also assessing the area occupied by atoms deviating from their predicted positions, the compactness of these identified regions should be considered a valuable metric for rewards. Given that compactness is a critical characteristic, the second component of the reward should focus on regions with minimal perimeter. By integrating both compactness and perimeter as objectives in our analysis, we establish a workflow that is both practical and dependable.

As depicted in **Figure 4(D)**, a set of optimal solutions was identified, demonstrating that no objective can be enhanced without adversely affecting another. By employing metrics to pinpoint the "best" solutions on the Pareto Frontier, the analysis effectively determined the optimal threshold and covariance type for Gaussian Mixture Model (GMM) clustering, as presented in **Figure 4(E)**. The deployment of the clustering map on the image of the YBCO substrate, as demonstrated in **Figure 4(F)**, effectively reveals areas within the YBCO structure where there is a higher likelihood of atoms deviating from their predicted positions.

To summarize, here we introduce an approach for the development of complex image analysis workflows based on the introduction of a reward function aligned with experimental objectives. This reward function is a measure of the success of analysis, and can be built based on simple physical consideration, comparisons to the oracle functions, or any other approach imitating human perception. With the reward function being defined, the image analysis problem reduces to that of the optimization in the combinatorial space of image operations and corresponding hyper-parameters, taking advantage of the immense volume of knowledge in his field.

Here, this approach has proven to be effective in a case study involving in situ ion irradiated $YBa_2Cu_3O_{7-\delta}$ layer images, where it facilitated the accurate identification of atomic positions and detection of amorphous regions. We propose the physics-based multi-objective reward functions for finding atom positions and classification of the amorphous regions and demonstrate the Bayesian optimization in the parameter space of multi-step simple image analysis functions to yield robust identification.

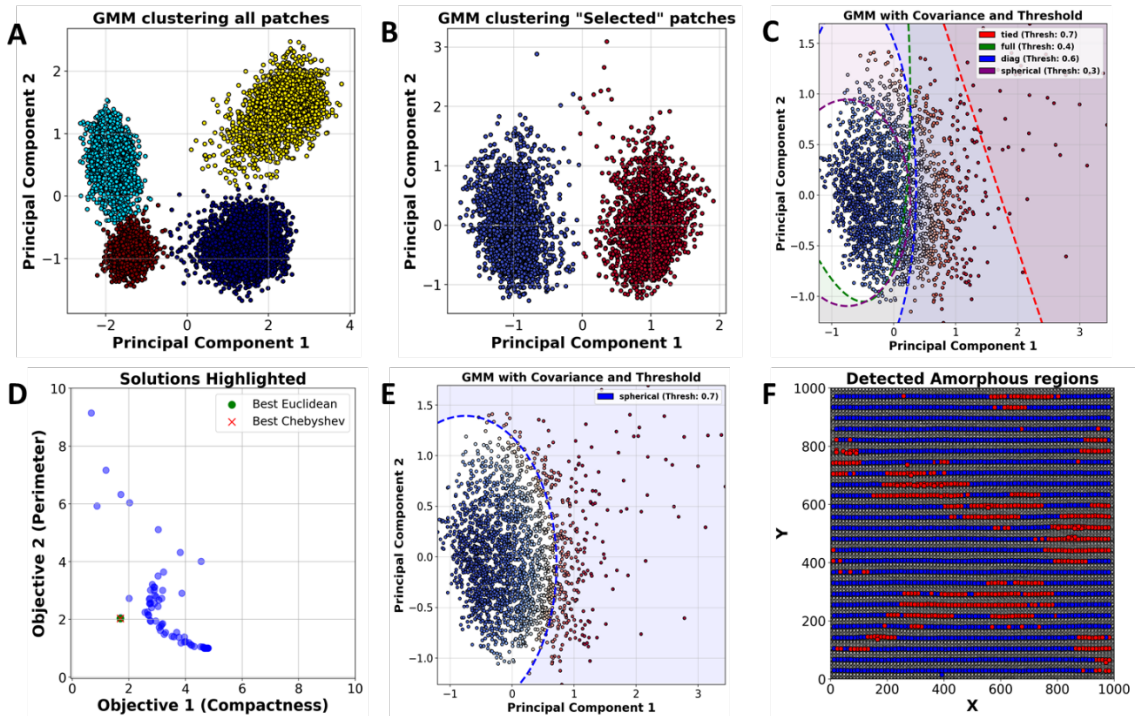

**Figure 4:** A) GMM clusters based on all the patches, providing 4 clusters with respect to 4 types of strong atoms in the YBCO structure. B) GMM clusters based on the patches centered on Ba atoms, presenting two types of Ba in the YBCO structure, C) GMM clusters based on only one type of Ba atoms, introducing some variety, which it can be differentiated by different values of threshold and covariance type in GMM clustering, D) Pareto Frontier solutions with respect to reward possession), E) Optimal threshold and covariance type achieved by MOBO for GMM clustering, and F) Uncovered amorphous areas in the substrate.

We believe that this approach has three significant impacts on microscopy. First, the introduction of a reward-function-based optimization approach makes the construction of analysis pipelines automated and unbiased, taking advantage of the powerful optimization approaches available today. Secondly, these analyses can be implemented as a part of automated experiments and real-time data analytics. Thirdly, the integration of reward functions across the domains offers a far more efficient approach for community integration than creation of disparate experimental data databases, contributing to the development of the open and FAIR experimental community.

**ACKNOWLEDGMENTS**:

This work (workflow development, reward-driven concept) was supported (K.B., S.V.K.) by the U.S. Department of Energy, Office of Science, Office of Basic Energy Sciences as part of the Energy Frontier Research Centers program: CSSAS-The Center for the Science of Synthesis Across Scales under award number DE-SC0019288. The authors (A.G. and H.Y.) acknowledge support from the DOE/EERE under contract No: DE-EE0007870.

**AUTHOR DECLARATIONS**

Conflict of Interest: The authors have no conflicts to disclose.

**Author Contributions:**

Kamyar Barakati: Formal analysis (equal); Writing – original draft (equal). Sergei V. Kalinin: Conceptualization (equal); Formal analysis (equal); Funding acquisition (equal); Supervision (equal). Hui Yuan: Investigation (equal). Amit Goyal: Investigation (equal).

**DATA AVAILABILITY:**

The code supporting the findings of this study is publicly accessible on GitHub at [**GitHub**]**.**